\documentclass[preprint]{emulateapj}       
\usepackage{amsmath,amssymb}
\usepackage{epsfig}     
\usepackage{color}
\newcommand{\hMsol}{{\>h^{-1}\rm M}_\odot}
\newcommand{\hMpc}{{\>h^{-1}\rm  Mpc}} 
\newcommand{\hkpc}{{\>h^{-1}\rm kpc}} 

\begin{document}
%---------------------------------------------------------------
\title{Assembly bias and the dynamical structure of dark matter halos}
%---------------------------------------------------------------
%
\author {  Andreas  Faltenbacher\altaffilmark{1,2,3} 
  and Simon D. M. White\altaffilmark{1}} 

\altaffiltext{1}{ Max Planck Institut f\"ur Astrophysik, 
  Karl-Schwarzschild-Str.  1, 85741 Garching,  Germany} 
\altaffiltext{2}{  MPA/SHAO  Joint Center  for Astrophysical 
  Cosmology at Shanghai Astronomical Observatory, Nandan
  Road  80,   Shanghai  200030,  China}  
\altaffiltext{3}{  Physics Department, University  of the Western  
  Cape, Cape Town  7535, South Africa}
\begin{abstract}
Based on  the Millennium Simulation  we examine assembly bias  for the
halo properties  : shape,  triaxiality, concentration, spin,  shape of
the velocity  ellipsoid and  velocity anisotropy.  For  consistency we
determine all these properties using the same set of particles, namely
all  gravitationally  self--bound  particles  belonging  to  the  most
massive  sub--structure  of  a  given  friends--of--friends  halo.  We
confirm  that  near--spherical  and  high--spin  halos  show  enhanced
clustering.   The  opposite  is   true  for  strongly  aspherical  and
low--spin  halos.  Further,  below the  typical collapse  mass, $M_*$,
more  concentrated   halos  show  stronger   clustering  whereas  less
concentrated  halos are less  clustered which  is reversed  for masses
above $M_*$. Going beyond earlier  work we show that: (1) oblate halos
are more strongly  clustered than prolate ones; (2)  the dependence of
clustering on the shape of  the velocity ellipsoid coincides with that
of the real--space  shape, although the signal is  stronger; (3) halos
with  weak velocity  anisotropy are  more clustered,  whereas radially
anisotropic  halos  are more  weakly  clustered;  (4)  for all  highly
clustered subsets we find systematically less radially biased velocity
anisotropy  profiles.   These  findings  indicate  that  the  velocity
structure of halos is tightly correlated with environment.
\end{abstract}
\keywords{cosmology:  theory --  cosmology :  dark matter  -- methods:
  N-body simulations}
\section{Introduction}
The dependence of halo clustering on a second parameter in addition to
mass is now  generally referred to as assembly  bias.  Analytic models
predict  the  clustering  of  halos  to depend  on  their  mass  alone
\citep{Kaiser-84,Cole-Kaiser-89,Mo-White-96}.                  However,
\cite{Gao-Springel-White-05}  and  various  subsequent studies  showed
that clustering  also depends on other halo  properties like formation
time,   concentration,   substructure    content,   spin   and   shape
\citep{Harker-06, Wechsler-06, Bett-07, Gao-White-07, Jing-Suto-Mo-07,
  Maccio-07a, Wetzel-07, Angulo-Baugh-Lacey-08}.

It soon became clear that  the dependency of clustering on the various
parameters  does  not simply  follow  from  the  relation among  those
parameters.   Here are  two examples:  (1) Earlier  formed  halos have
higher      concentrations     \citep[e.g.,][]{Navarro-Frenk-White-97,
  Wechsler-02, Neto-07}  nevertheless the dependence  of clustering on
formation   time    and   concentration   show    different   behavior
\citep[cf.,][]{Jing-Suto-Mo-07}.   (2) As discussed  in \cite{Bett-07}
there is  a weak  correlation between spin  and shape,  more spherical
halos show on  average a slightly lower spin  parameter.  However, the
bias behavior is  opposite to what one would  naively derive from this
correlation.   The most  nearly  spherical halos  and  the halos  with
highest   spins   are  clustered   above   average  \citep[see   also,
][]{Gao-White-07}.  These examples  suggest that the ranking according
to any  given parameter, like  concentration, spin or shape  cannot be
converted in  any simple way  to the approximate ranking  according to
another   parameter.   In   this  context   \cite{Croton-Gao-White-07}
speculated  (but  did  not  demonstrate)  that there  may  be  a  more
fundamental parameter capable of uniquely predicting halo clustering.

Simple implementations  of extended Press-Schechter  and excursion set
models  \citep{Press-Schechter-74,Bond-91,Lacey-Cole-93} commonly used
to estimate halo  statistics do not predict a  dependence of formation
time on  environment.  However, using  a mass filter  in configuration
space  rather  than in  k-space  \cite{Zentner-07} demonstrated  that
halos in denser  environments do form later independent  of halo mass.
At the high mass end this agrees with findings from N-body simulations
\citep[e.g.,][]{Wechsler-02,Jing-Suto-Mo-07}  but is  opposite  to the
behavior observed at the low mass end \citep[e.g.,][]{Gao-Springel-White-05}.

\cite{Dalal-08}  argued that  for high  halo masses  assembly  bias is
expected  from  the  statistics   of  the  peaks  of  Gaussian  random
fluctuations  and at  low masses  it  arises from  a subpopulation  of
low-mass halos whose mass accretion has ceased.  It is unclear whether
whether other aspects  of assembly bias can be  explained in this way.

Several  other  studies  have  investigated  the  dependence  of  halo
formation   times   or,  similarly,   merger   rates  on   environment
\citep[e.g.,][]{Gottloeber-Klypin-Kravtsov-01,           Gottloeber-02,
  Sheth-Tormen-04, FakhouriMa-09a, FakhouriMa-09b, Hahn-09}. Although,
slightly different  density estimators, like over density  in a sphere
or  mark correlation functions,  are employed  it is  generally agreed
upon that halos less  massive than $\sim10^{13}\hMsol$ which reside in
high density  regions form earlier  compared to those with  equal mass
but located in less dens regions.

Most of the previous studies  examined the impact of the assembly bias
on galaxy clustering  statistics. This is not the  prime focus of this
work.  We are mainly  concerned about the interplay between clustering
and  the dynamical  structure  of halos.   To  describe the  dynamical
structure of  halos we use a  variety of parameters,  namely: the mean
shape, the  mean triaxiality,  the spin, the  non-dimensionlized total
velocity  dispersion (a measure  of concentration),  the shape  of the
global  velocity  dispersion tensor,  and  the mean  radial/tangential
velocity anisotropy.   We also explore whether subsets  of halos which
show assembly bias share any common property. The most obvious we find
is  the  behavior of  velocity  anisotropy.   On  average, all  highly
clustered  subsets  appear  to  have  less  radially  biased  internal
motions.  The opposite holds for less clustered subsets.
\begin{figure*}
  \epsfig{file=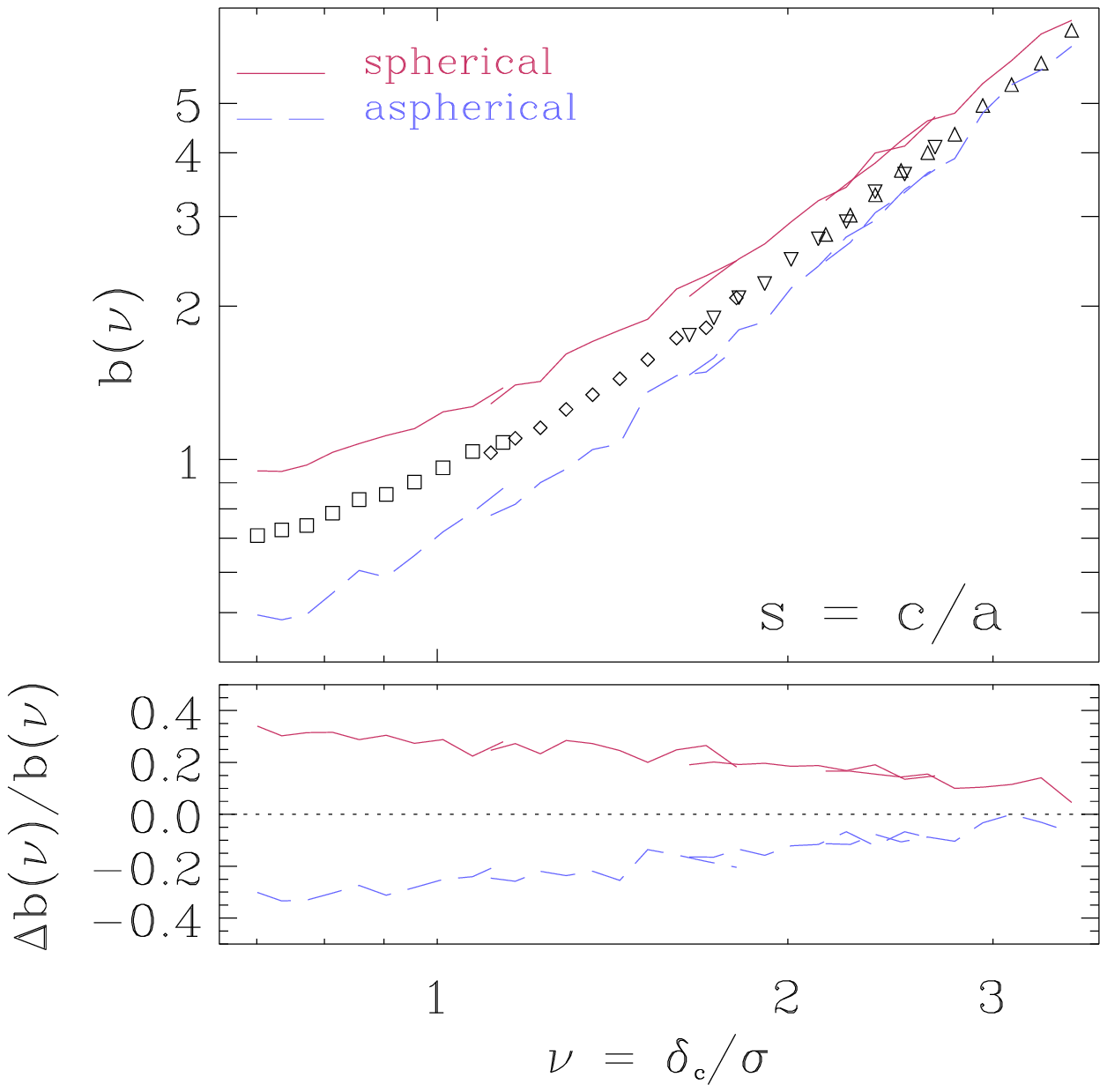,width=0.315\hsize}
  \epsfig{file=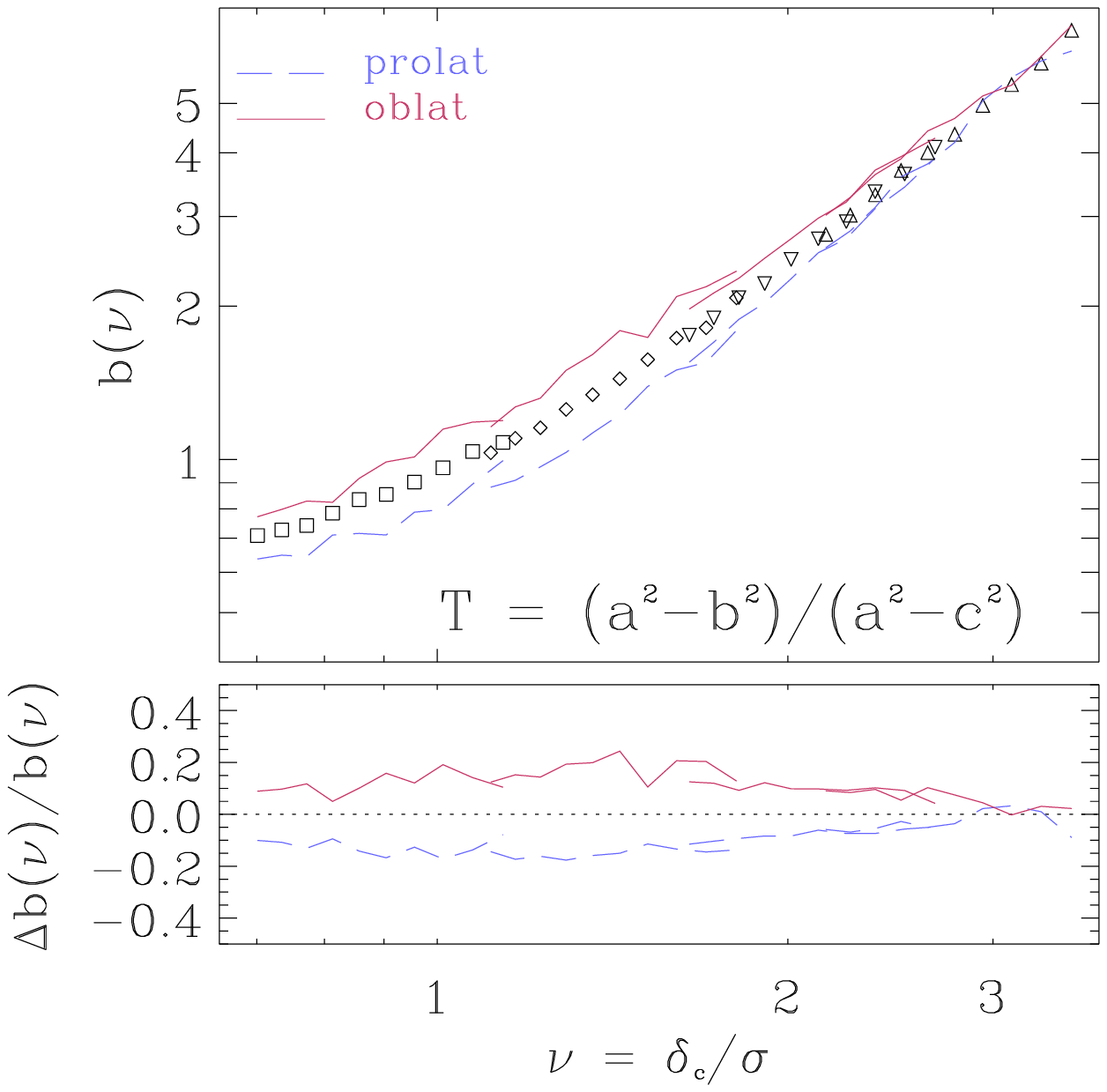,width=0.315\hsize}
  \epsfig{file=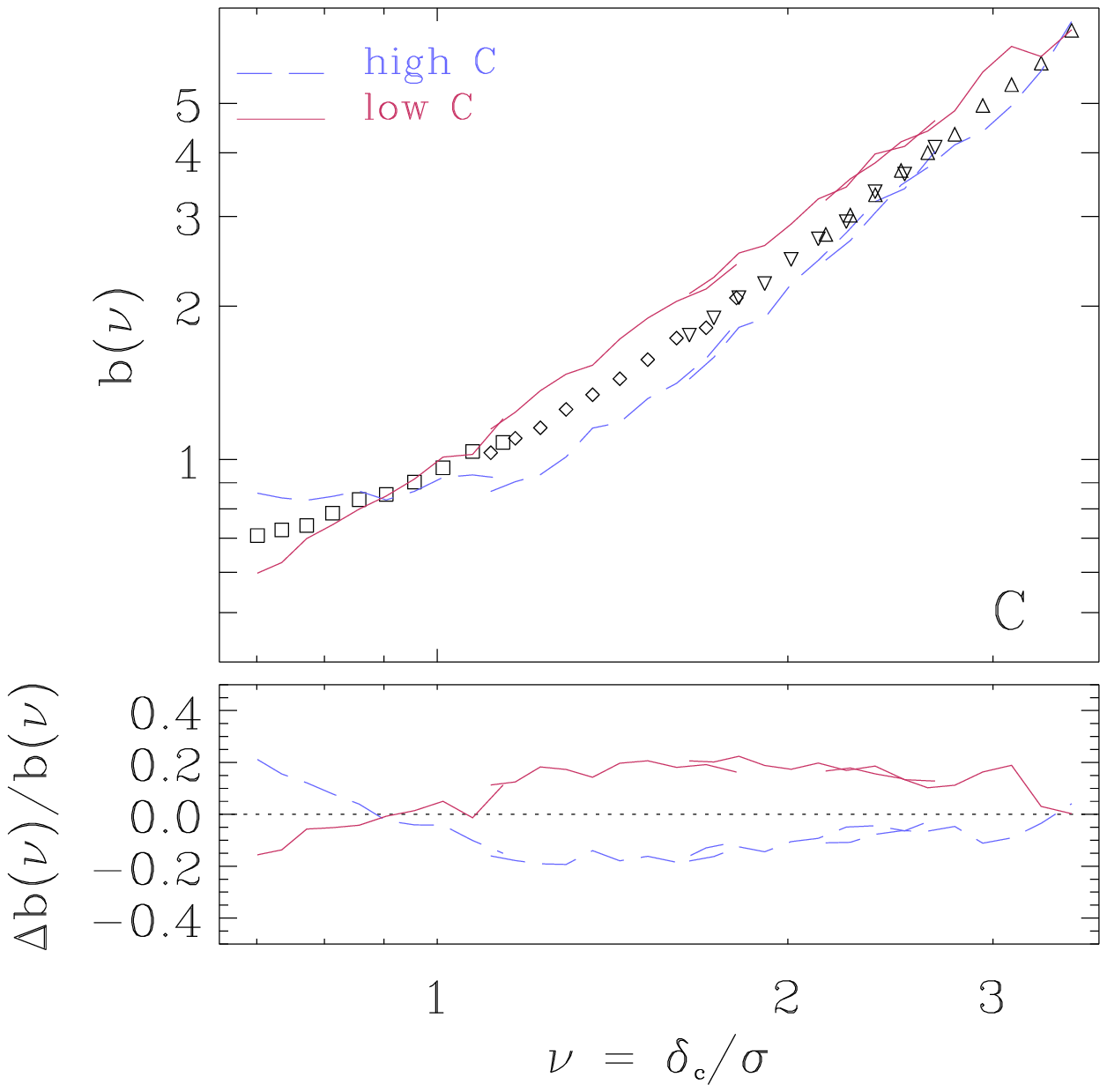,width=0.315\hsize}\\
  \epsfig{file=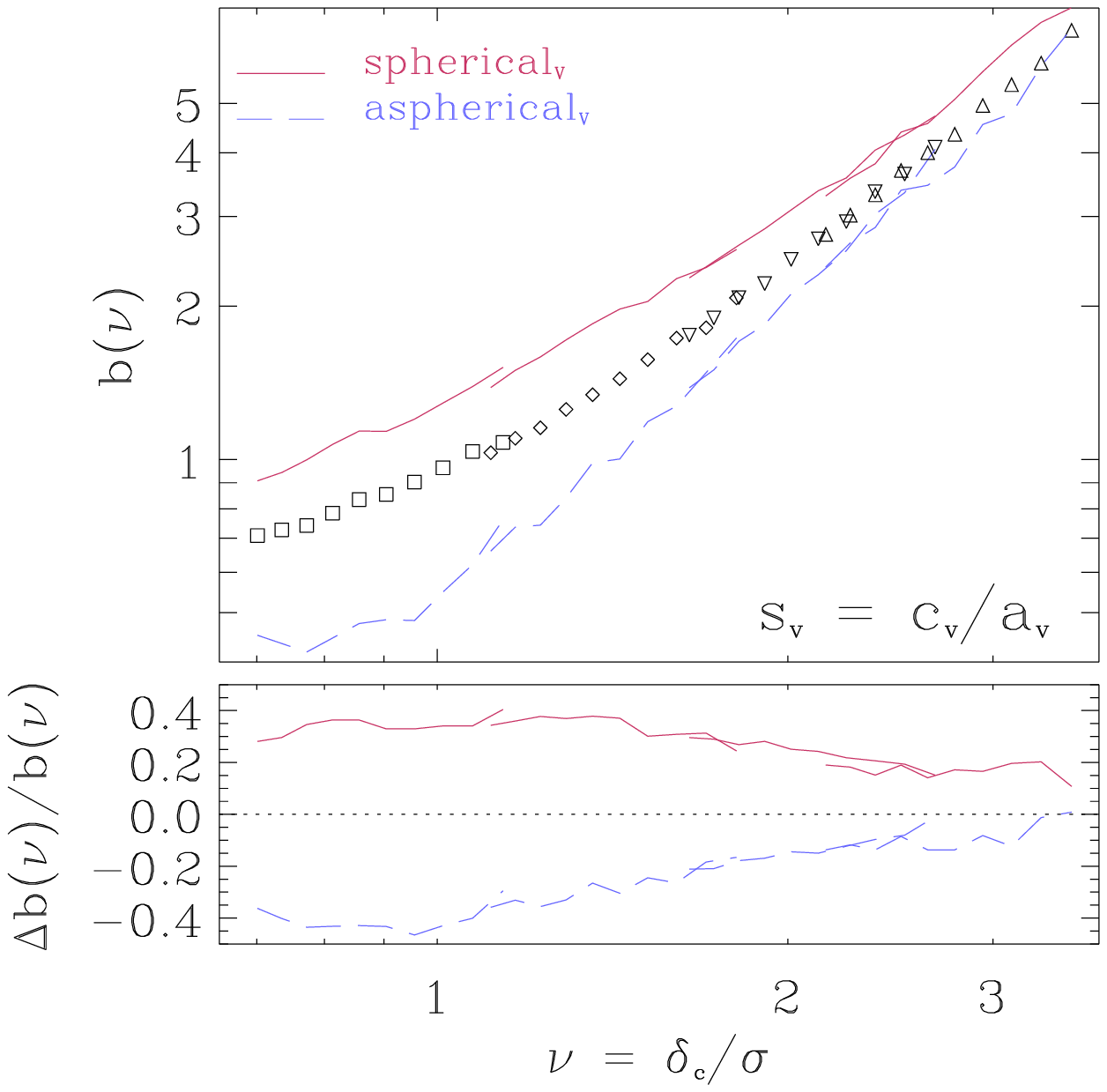,width=0.315\hsize}
  \epsfig{file=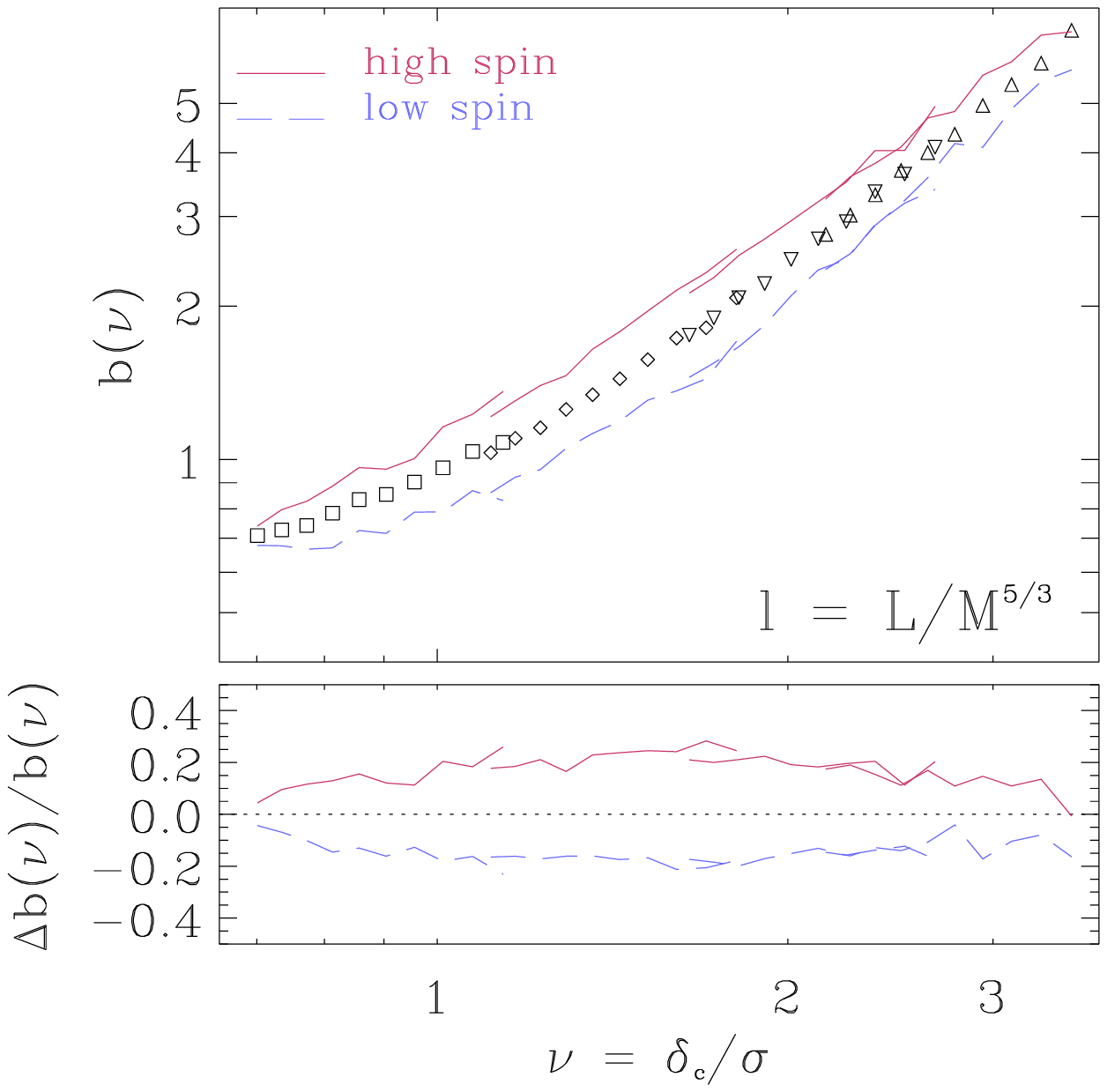,width=0.315\hsize}
  \epsfig{file=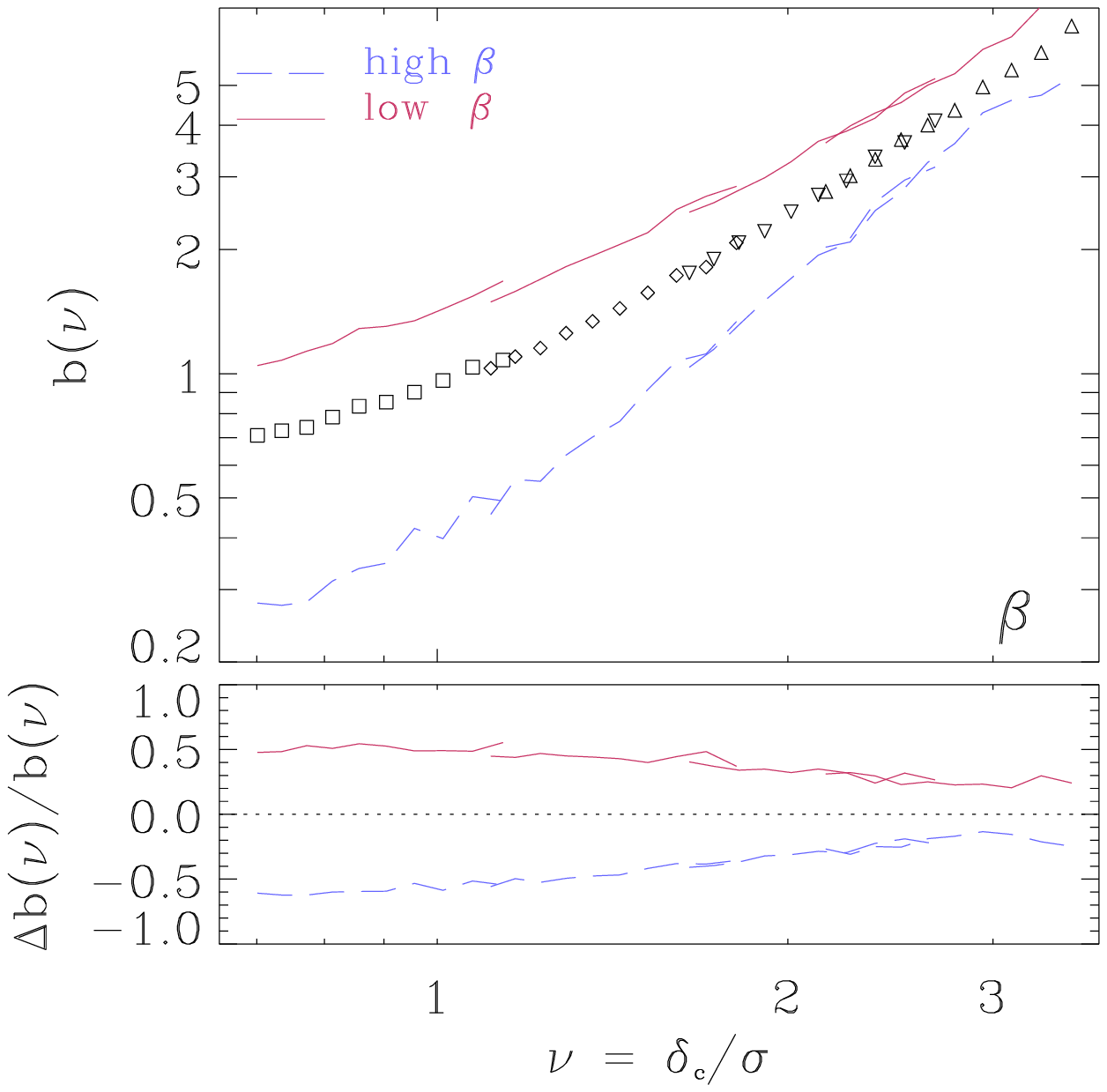,width=0.315\hsize}
  \caption{\label{fig:D} Bias  factor as a  function of halo  mass and
    halo  properties. Halo  mass is  given parametrically  through the
    equivalent  peak  height  $\nu=\delta_c/\sigma$.   The  additional
    properties in the six panels are: shape ($s$), Triaxiality ($T$) ,
    concentration ($C$), shape  of the  velocity ellipsoid
    ($s_{\rm v}$),  the  spin   ($L$)  and  the  velocity  anisotropy
    ($\beta$).  Symbols are  bias factors defined for all  haloes in a
    given mass bin.  The  various shapes (squares, diamonds, two kinds
    of triangles)  indicate the  redshifts (0,1,2,3) of  the snapshots
    used  to determine  them.  The  same  redshifts are  used for  the
    computation of the solid curves  which are bias factors for haloes
    in the lower  and upper 20 \% tails  of the distributions of
    the particular property concerned.}
\end{figure*}
\begin{figure*}
  \epsfig{file=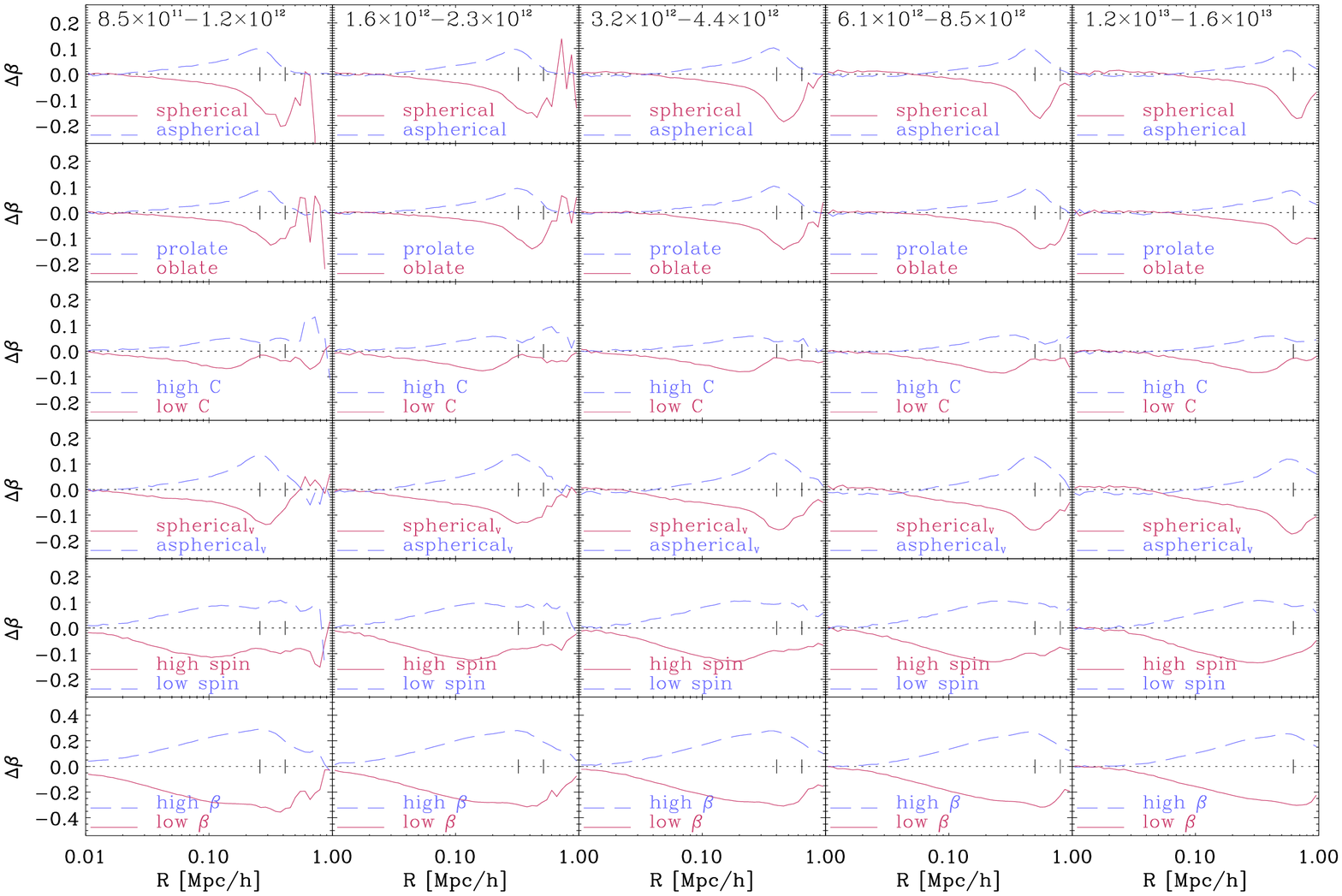,width=0.945\hsize}
  \caption{\label{fig:dbeta}  Difference between the  average velocity
    anisotropy profiles,  $\Delta\beta$, of halos  in the 20  per cent
    tails of the according distribution and all halos at a given mass.
    The same halo properties are used as in Fig.~\ref{fig:D}. From top
    to   bottom  they   are:  shape,   triaxiality,   scaled  velocity
    dispersion,  shape  of   velocity  ellipsoid,  spin  and  velocity
    anisotropy.   Columns  correspond   to  every  second  symbol  for
    $\nu\lesssim 1$ in Fig.~\ref{fig:D}.  The two short vertical lines
    indicate the averages of the radii including a mean density of 200
    times  the  critical  and  200  times  the  cosmic  mean  density,
    respectively. The averages are based  on all halos in a given mass
    bin.  Averages  based on halos  in the tails of  the distributions
    show only marginal differences.  In the rightmost column the outer
    marks  exceed  the  plot  range.   Signals beyond  the  marks  are
    dominated by effects of the boundaries of our halos.}
\end{figure*}
\section{Methodology}
\label{sec:data}
\subsection{Simulation}
The  Millennium  Simulation  \citep{Springel-05a} adopted  concordance
values  for  the parameters  of  a  flat  $\Lambda$ cold  dark  matter
($\Lambda$CDM)  cosmological  model,  $\Omega_{\rm  dm} =  0.205$  and
$\Omega_{\rm b} = 0.045$ for the current densities in CDM and baryons,
$h = 0.73$ for the present dimensionless value of the Hubble constant,
$\sigma_8 =  0.9$ for the rms  linear mass fluctuation in  a sphere of
radius $8\hMpc$ extrapolated to $z = 0$,  and $n = 1$ for the slope of
the primordial fluctuation spectrum.  The simulation followed $2160^3$
dark matter particles from $z= 127$  to the present day within a cubic
region $500\hMpc$ on a side resulting in individual particle masses of
$8.6\times10^8\hMsol$.      The    gravitational    force     had    a
Plummer-equivalent comoving softening of $5\hkpc$.  The Tree-PM N-body
code GADGET2 \citep{Springel-05b} was used to carry out the simulation
and the full data were stored 64 times spaced approximately equally in
the logarithm of the expansion factor.
\subsection{Halo sample and properties}
\label{sec:halo}
The halos  are found by a  two-step procedure.  In the  first step all
collapsed  halos with  at least  20 particles  are identified  using a
friends-of-friends (FoF) group-finder with  linking parameter b = 0.2.
These objects will be  referred to as FoF-halos.  Then post-processing
with         the         substructure        algorithm         SUBFIND
\citep{Springel-01} subdivides each  FoF-halo into a set
of  self-bound  sub-halos. Here  we  only  consider  the most  massive
sub-halo within  each FoF--halo which on  average comprises $\sim80\%$
of  the total  mass of  the FoF--halo.   We refer  to  these sub-halos
simply as  halos. As is common  practice we chose the  position of the
most bound particle of that sub--halo as center of the halo.  All halo
properties are  consistently computed  based on particles  assigned to
that  (most  massive sub-)  halo.   That  way  possible confusion  due
properties which are computed  based on different halo definitions can
be excluded.  For instance,  commonly the spin parameter is determined
withn a  spherical volume whereas the  shape is based  on all particles
associated with the FoF--halo. In this study, however, both quantities
are computed based on the same set of particles.

As    is   standard,    the   shape,    $s=a/c$,    the   triaxiality,
$T=(a^2-b^2)/(a^2-c^2)$,  and the  shape of  the  velocity ellipsoids,
$s_{\rm v}=a_{\rm v}/c_{\rm v}$  are computed based on the eigenvalues
($a_{\rm(v)}>b_{\rm(v)}>c_{\rm(v)}$)  of  the  second moments  of  the
spatial and velocity distributions of all halo particles. The velocity
anisotropy   parameter   is   given   by   $\beta=1-0.5\,\sigma^2_{\rm
  t}/\sigma^2_{\rm  r}$, where $\sigma_{\rm  r}$ and  $\sigma_{\rm t}$
are the  radial and tangential  velocity dispersion based on  all halo
particles. Since the conventional  determination of the spin parameter
and the concentration is based on the particle content within a sphere
and  not  on  self-bound  particles we  introduce  slightly  different
quantities  to  replace them.   The  spin,  $l$,  is computed  as  the
absolute value of the total  angular momentum multiplied by the Hubble
constant $H$ to the one third  power and divided by the total mass $M$
to the  power $5/3$. This is  closely related to the  formula given in
\cite{Bullock-01}.  As  quantity measuring concentration, we  use $C =
\sigma/(H M)^{1/3}$ where $\sigma$ is the total velocity dispersion of
the halo.  Concentration and  $\sigma$ are positively correlated, i.e.
high $\sigma$ values correspond to high concentrations and vice versa.
The powers  of mass in the denominators  of $l$ and $C$  are chosen to
compensate for the intrinsic mass dependence of these quantities.
\subsection{Determining the bias factor} 
The determination of the bias  factor, $b$, follows the approach given
in \cite{Gao-White-07}.  It is  computed as the relative normalization
factor  which  minimizes  the  mean  square of  the  difference  $\log
\xi_{\rm hm}- \log  b\xi_{\rm mm}$ for four equal  width bins in $\log
r$ spanning  the comoving separation range  $6 < r <  20 \hMpc$. Here,
$\xi_{\rm  hm}$  is  the  halo  mass  cross-correlation  function  and
$\xi_{\rm  mm}$ is  the  mass autocorrelation  function.  Compared  to
estimators based  on the halo autocorrelation  function this estimator
has highly improved noise  characteristics because of the large number
of  dark matter particles  available.  As  suggested by  standard halo
bias   models   \citep{Mo-White-96}   and  verified   numerically   by
\cite{Gao-White-07} this approach is equivalent to estimators based on
halo autocorrelation functions.  The  large-scale bias depends on mass
and  redshift   through  the   equivalent  peak  height   $\nu(M,z)  =
\delta_c(z)/\sigma(M,z)$,  where  $\sigma(M,z)$   is  the  rms  linear
overdensity within a  sphere which contains the mass  $M$ in the mean,
and $\delta_c(z)$ is the  linear overdensity threshold for collapse at
redshift  $z$.   The  typical  collapse mass,  $M_*$,  corresponds  to
$\nu=1$. We will use the  equivalent peak height to seamlessly combine
results derived at redshifts $z=0,1,2$ and $3$.
\section{Results}
\label{sec:results}
\subsection{Assembly bias based on consistent halo properties}
\label{sec:gao}
Fig.~\ref{fig:D} shows  the assembly bias based on  halo properties as
defined in  \$~\ref{sec:halo}. Symbols (squares,  diamonds, triangles)
display bias factors  for all halos.  Solid and  dashed lines give the
bias based on the upper and  lower 20 \% tiles of the distributions of
each  specific property  as indicated  by  labels in  the lower  right
corners.  We use solid red lines for the more strongly biased tail and
dashed blue lines for the  more weakly biased tail.  These line styles
are  adopted in  Fig~\ref{fig:dbeta} to  ease comparison.   Given that
$C$ and $l$ correspond  closely to conventional concentration and
spin we confirm the results  reported in previous work.  However, with
triaxiality, $T$,  shape of the  velocity ellipsoid, $s_{\rm  v}$, and
the velocity  anisotropy, $\beta$, we add three  more properties which
show significant assembly bias.

The displayed dependency of bias on shape, $s$, reproduces the results
by  \cite{Bett-07}. For  all masses,  more nearly  spherical  halos are
clustered  above, and  strongly aspherical  halos below  average.  The
behavior  of the triaxiality,  $T$, indicates  that prolate  halos are
clustered below  average and oblate halos are  more strongly clustered
than  average. The results  for the  scaled velocity  dispersion, $C$,
which here  replaces the conventional concentration,  indicate that for
$\nu\lesssim1$ high concentration halos are more and low concentration
halos are less clustered.  This behavior reverses sign for $\nu\gtrsim
1$.   The  same  trends  have  been  reported  in  \cite{Wechsler-06},
\cite{Gao-White-07}, \cite{Jing-Suto-Mo-07} and \cite{Wetzel-07}.  The
particularly good agreement with Fig.2 in \cite{Jing-Suto-Mo-07} based
on the  conventional concentration  increases confidence that  $C$ can
represent concentration  in the context  of assembly bias  studies.  The
shape of the velocity ellipsoid, $s_{\rm v}$, behaves similarly to the
shape of the spatial  distribution, $s$.  The dependence of clustering
on  spin, $l$,  is comparable  to  the results  in \cite{Bett-07}  and
\cite{Gao-White-07}.   Finally, halos  with  low velocity  anisotropy,
$\beta$,  are more  clustered and  the opposite  holds for  halos with
strongly radially  biased velocities.  The signal based  on $\beta$ is
the  most pronounced  among all  shown here  indicating  that velocity
related  quantities   are  tightly  correlated   with  the  clustering
strength.

We also obtained results for the dependence of clustering on formation
time  (not  shown) as  examined  in \cite{Gao-Springel-White-05}.   As
formation time for their {\it  fof-halos} they used the time when half
the final was acquired.  For  a variety of reasons we here restrict
ourselves to  self--bound main subhalos  which are simpler  and better
defined objects.  Thus we computed the formation times in the same way
as \cite{Gao-Springel-White-05} except that  we use main subhalos.  It
turned  out that  the effect  is qualitatively  similar to  that found
earlier but  is {\it much}  weaker. We find  only a $\sim10$  per cent
increase  of  the  bias  factor  for  early  formed  halos  at  masses
corresponding to $\nu\lesssim  1$. 
Different factors like the orbital parameters or the concentrations of
the merging  systems imply a  certain stochasticity of the sub-halo 
masses  which  propagates  into  the  determination  of  the formation 
times and reduces the assembly bias signal.
\cite{Li-Mo-Gao-08} noted  that the  definition of the  formation time
has  a  substantial impact  on  the  strength  of the  assembly  bias,
confirming that  the change from fof- to  sub-halo accretion histories
is responsible for the reduction of the assembly bias signal.
\subsection{Do highly clustered halo subsets share similar properties?}
\label{sec:fundamental}
To  find  out whether  highly  clustered  halo  subsets share  similar
properties we compute  stacked profiles of the halos  belonging the 20
\%  tail subsets of a  given property, and compare  them with the
stacked profiles of all halos.   We examine the difference in density,
velocity  dispersion,  phase  space  density and  velocity  anisotropy
profiles.  Density  and velocity  dispersion profiles show  no unique
trend with the bias behavior.

The average phase space density profiles determined by $\rho/\sigma^3$
indicate  that more  strongly clustered  subsets, solid  red  lines in
Fig.~\ref{fig:D}, tend to lie above  the average profiles based on all
halos in  a given mass bin.   However, for some  properties, like spin
and velocity  anisotropy, the differences  are small and  dependent on
radius.

Motivated      by       studies      of      \cite{Hansen-06}      and
\cite{Schmidt-Hansen-Maccio-08} we  also probe the  behavior for phase
space   densities   computed   as   $\rho/\sigma_{\rm   r}^3$,   where
$\sigma_{\rm r}$  denotes the radial velocity  dispersion.  For masses
larger than  $M_*$ the average  phase space profiles of  subsets which
show  stronger  clustering   (solid  red  lines  in  Fig.~\ref{fig:D})
systematically lie above the  average profiles of all halos.  Profiles
based on less clustered subsets  fall below the average profiles based
on all  halos in  a given  mass bin.  However,  for halo  masses below
$M_*$ the reversal of sign seen in concentration (velocity dispersion)
is not reproduced.   \cite{Schmidt-Hansen-Maccio-08} argued that there
is      no       empirical      justification      \citep[but      see
  also][]{Dehnen-McLaughlin-05} for the particular definition of phase
space density used here, and we find that changing the exponent of the
velocity  dispersion  definitely  has  an impact  on  the  correlation
between   clustering  and   phase   space  profiles.    Due  to   such
uncertainties we leave the discussion at this point.  Nevertheless, we
note  there is a  systematic trend  for more  clustered halos  to have
higher phase space densities.

In  contrast  to the  phase  space  profiles  the velocity  anisotropy
profiles, $\beta(r)=1-0.5\sigma^2_{\rm r}(r)/\sigma^2_{\rm t}(r)$, are
unambiguously  defined ($\sigma_{\rm  r}(r)$  and $\sigma_{\rm  t}(r)$
denotes the radial and the tangential velocity dispersions averaged in
spherical   shells).   Fig.~\ref{fig:dbeta}   shows   the  difference,
$\Delta\beta$, between the stacked velocity anisotropy profiles of all
halos and the halos belonging to  20 \% tails of the distribution of a
given  property.  These  properties are,  from top  to  bottom: shape,
triaxiality,  scaled  velocity   dispersion,  shape  of  the  velocity
ellipsoid,  spin  and total  velocity  anisotropy.   The columns  show
various  mass  bins  corresponding  to  every  second  mass  bin  with
$\nu\lesssim1$    in    Fig~\ref{fig:D}.     The    comparison    with
Fig.~\ref{fig:D}  indicates  that more  clustered  subsets (solid  red
lines) show on average lower $\beta$ profiles.  The opposite holds for
the  less  clustered subsets  (dashed  blue  lines).   For the  scaled
velocity dispersion, $C$, the  difference is strongly reduced, however
the  reversal of  the  sign  as apparent  in  Fig.~\ref{fig:D} is  not
explicitly  reproduced.   But  apart  from  that,  there  is  a  clear
correlation between clustering behavior and velocity anisotropy.

One possible explanation for the more nearly isotropic global velocity
distribution of more clustered halos  is that the impact parameters of
the  merging sub-halos  are  larger due  to gravitational  deflections
short  time before  accretion. \cite{FakhouriMa-09b}  have  shown that
mergers are  the dominant  mode of accretion  in high  density regions
thus  they  presumably  determine  the  velocity  structure  of  these
halos. On  the other side  in lower density regions  the gravitational
field is dominated  by the halo itself and accretion  occurs in a more
radial fashion leaving the  observed imprint on the velocity structure
of the halo.
\section{Conclusion}
\label{sec:conclusion}
Taking advantage  of the large  volume and the superior  resolution of
the Millennium  Simulation we reassess assembly bias  for various halo
properties. We conclude with a recapitulation of the main results:

1) Based on consistent determinations of various halo properties, like
shape,  spin and  concentration, we  confirm the  results  reported in
previous  studies.  More  nearly  spherical and  high--spin halos  are
clustered above and aspherical and  low--spin below average.  Below the
typical collapse  mass, $M_*$,  more concentrated halos  show stronger
clustering whereas  less concentrated  halos are less  clustered. This
reverses  for masses  above $M_*$.   The clustering  cannot  simply be
explained by the relations among the properties.

2) To the halo properties which  have already been shown to correlate with 
clustering behavior, we  have added the triaxiality, the  shape of the
global  velocity  ellipsoid  and  the velocity  anisotropy  parameter.
Oblate  halos, halos  with  a more  nearly  isotropic global  velocity
distribution  and  halos  with   weak  velocity  anisotropy  are  more
clustered than  average.  Contrary properties  reduce clustering. Very
prominent  signals  are found  for  the velocity--related  properties,
indicating that halo velocity structure is tightly correlated with the
clustering behavior.

3)  We also  showed  that  independent of  the  halo property  (shape,
triaxiality, spin,  concentration) the velocity  anisotropy profile of
the  more  clustered subsets  lies  systematically  below the  average
profile, whereas less clustered subsets show profiles, indicating more
radially biased internal motions.  

Our findings  show that  the internal velocity  structure of  halos is
strongly  influenced by  environment.  Halos  are not  "universal" and
their  internal  properties  (the  deviations from  universality)  are
related in complex ways to their environment.
\section*{Acknowledgements} 
We  thank  the  anonymous  referee  for helpful  suggestions.   AF  is
grateful for kind hospitality  of SAAO.  The Millennium Simulation was
carried out  by the Virgo  Consortium on the Regatta  supercomputer of
the Computing Centre of  the Max-Planck-Society in Garching.  Data for
halos     and      galaxies     are     publicly      available     at
http://www.mpa-garching.mpg.de/millennium.
%
%\bibliography{lit}

%
\end{document}